\documentclass[12pt,a4paper]{article}
\pdfoutput=1

\usepackage{amsmath,amssymb,bbm} 
\usepackage{accents}
\usepackage[usenames,dvipsnames]{color}
\usepackage[bookmarks=true,hyperfigures=true]{hyperref}
\usepackage{graphicx}
\usepackage[nosort]{cite}
\usepackage[bulletsep]{collref}
\usepackage{tensor}

\usepackage[a4paper,text={173mm,216mm},centering]{geometry}

\allowdisplaybreaks[3]

\usepackage[font=small,labelfont=bf,width=0.85\textwidth]{caption}

\let\oldbfseries=\bfseries
\let\oldmdseries=\mdseries
\let\oldnormalfont=\normalfont
\renewcommand{\bfseries}{\oldbfseries\boldmath}
\renewcommand{\mdseries}{\oldmdseries\unboldmath}
\renewcommand{\normalfont}{\oldnormalfont\unboldmath}

\makeatletter
\newlength{\apb@width}
\newcommand{\autoparbox}[2][c]{\settowidth{\apb@width}{#2}\parbox[#1]{\apb@width}{#2}}

\makeatother

\newcommand{\nn}{\nonumber}

\newcommand{\be}{\begin{equation}}
\newcommand{\ee}{\end{equation}}
\newcommand{\ba}{\begin{eqnarray}}
\newcommand{\ea}{\end{eqnarray}}

\ifx\genfrac\sdflkaj\else\fi

\newcommand{\ft}[2]{{\textstyle\frac{#1}{#2}}}

\newcommand{\cM}{\mathcal{M}}
\newcommand{\cO}{\mathcal{O}}

\def\l<{\langle}\def\r>{\rangle}

\makeatletter
\newcommand{\namedref}[2]{\hyperref[#2]{#1~\ref*{#2}}}
\newcommand{\secref}{\@ifstar{\namedref{Section}}{\namedref{sec.}}}
\newcommand{\subsecref}{\@ifstar{\namedref{Subsection}}{\namedref{subsec.}}}
\newcommand{\appref}{\@ifstar{\namedref{Appendix}}{\namedref{app.}}}
\newcommand{\tabref}{\@ifstar{\namedref{Table}}{\namedref{table}}}
\newcommand{\figref}{\@ifstar{\namedref{Figure}}{\namedref{figure}}}
\makeatother

\newif\ifmrnote 
\mrnotetrue
 
\usepackage[active]{srcltx}

\newif\ifjbnote 
\jbnotetrue

\newcommand{\vev}[1]{\langle#1\rangle}
\newcommand{\bev}[1]{ [#1]}

\newcommand{\eqn}[1]{(\ref{#1})}

\newcommand{\YM}{\text{YM}}
\newcommand{\GR}{\text{G}}

\newcommand{\se}{\epsilon}							  

\newcommand{\tree}{\text{\, tree}}
\newcommand{\overbarM}[1]{\mkern 4.5mu\overline{\mkern-4.5mu#1\mkern-0.0mu}\mkern 0.0mu}

\begin{document}
\thispagestyle{empty}

\begingroup\raggedleft\footnotesize\ttfamily
HU-EP-14/36\\
\vspace{5mm}
\endgroup

\begin{center}
{\LARGE\bfseries Factorized soft graviton theorems \\ at loop level
\par}%
\vspace{15mm}

\begingroup\scshape\large 
 Johannes Broedel${}^{1}$, Marius de Leeuw${}^{1,2}$, \\[0.3cm] Jan Plefka${}^{1,3}$ and
 Matteo Rosso${}^{1}$
\endgroup
\vspace{5mm}

\textit{${}^{1}$ Institut f\"ur Theoretische Physik,\\ Eidgen\"ossische
  Technische Hochschule
  Z\"urich,\\ Wolfgang-Pauli-Strasse 27, 8093 Z\"urich, Switzerland}\\[0.1cm]
  \texttt{\small \{jbroedel,mrosso\}@itp.phys.ethz.ch\phantom{\ldots}} \\ \vspace{5mm}

\textit{${}^{2}$ Niels Bohr Institute, \\ Copenhagen University,\\
Blegdamsvej 17, 2100 Copenhagen \O, Denmark}\\[0.1cm]
  \texttt{\small deleeuwm@nbi.ku.dk \phantom{\ldots}} \\ \vspace{5mm}

\textit{${}^{3}$ Institut f\"ur Physik und IRIS Adlershof,\\ Humboldt-Universit\"at zu Berlin, \phantom{$^\S$}\\
  Zum Gro{\ss}en Windkanal 6, 12489 Berlin, Germany} \\[0.1cm]
\texttt{\small plefka@physik.hu-berlin.de\phantom{\ldots}} \vspace{8mm}

\textbf{Abstract}\vspace{5mm}\par
\begin{minipage}{14.7cm}
We analyze the low-energy behavior of scattering amplitudes involving gravitons
at loop level in four dimensions. The single-graviton soft limit is controlled
by soft operators which have been argued to separate into a factorized piece
and a non-factorizing infrared divergent contribution. 
In this note we show that the soft operators responsible for the factorized
contributions are strongly constrained by gauge and Poincar\'e invariance under
the assumption of a local structure. 
We show that the leading and subleading orders in the soft-momentum expansion
can not receive radiative corrections. The first radiative correction occurs
for the sub-subleading soft graviton operator and is one-loop exact.  It
depends on only two undetermined coefficients which should reflect the field
content of the theory under consideration.

\end{minipage}\par
\end{center}
\newpage


\section{Introduction}

Recently the low energy behavior of graviton scattering amplitudes with a
single soft graviton momentum has been related
\cite{Strominger:2013jfa,He:2014laa,Kapec:2014opa} to Ward identities of the
extended Bondi, van der Burg, Metzner and Sachs (BMS) symmetry
\cite{Bondi:1962px,Sachs:1962wk,Barnich:2010eb}. The role of BMS symmetry as a
potential hidden symmetry of the quantum gravity S-matrix in asymptotically
flat four-dimensional space-time has triggered considerable interest.

In the soft limit the $(n+1)$-point scattering amplitude is dominated by a soft
pole as was shown by Weinberg about fifty years ago \cite{Weinberg:1964ew}. The
soft amplitude factorizes on the pole into a universal soft function and the
remaining hard graviton $n$-point amplitude. This property holds in any
spacetime dimension.

Revived by the work of Cachazo and Strominger \cite{Cachazo:2014fwa} the
universal factorization has been shown to extend to sub- and sub-subleading
order in the soft-momentum expansion
\cite{Schwab:2014xua,Afkhami-Jeddi:2014fia,Kalousios2014,Zlotnikov2014}.  In
distinction to the leading Weinberg pole, which is a function of the soft and
hard momenta and the soft graviton polarization, the sub- and sub-subleading
soft behavior of the $(n+1)$-point graviton amplitude can be expressed in terms
of differential operators in the hard momenta and hard graviton polarizations
acting on the hard $n$-point amplitude.

Gluon amplitudes exhibit a similar universal leading and subleading soft
behavior at tree level. Known as Low's theorem
\cite{Low:1958sn,Burnett:1967km}, the form of the soft operators was recently
recast into the language of modern on-shell techniques \cite{Casali:2014xpa}.
The soft limit of open-string tree amplitudes was studied in
refs.~\cite{Schwab:2014fia,Bianchi:2014gla}, where further corrections to the
field-theory soft theorems have been excluded. Interestingly, the subleading
soft theorems for gravitons and gluons may also be derived from the soft limit
of vertex operators in a dual ambitwistor string theory \cite{Geyer:2014lca}.

In a series of recent papers \cite{Bern:2014vva,White:2014qia} the existence of
subleading soft-graviton and soft-gluon theorems was shown to be a 
consequence of on-shell gauge invariance and Feynman diagrammatic reasoning.
Simultaneously, an alternative point of view was put forward by the present
authors in ref.~\cite{Broedel:2014fsa}: The tree-level soft-graviton and
soft-gluon expressions are composed from a highly restricted class of operators
compatible with on-shell gauge invariance, factorization and Poincar\'e
symmetry. 
This result could be established without referring to underlying Feynman
diagrammatics at all.  For the graviton case, this reasoning supplements the
established sub- and sub-subleading soft operators with one further candidate
each.

Given these results it is natural to ask whether the soft-graviton theorems
receive corrections at loop level \cite{Bern:2014oka,He:2014bga}. In
four-dimensional gravity the dimension of the coupling constant $\kappa$ leads
to strong constraints on the possible loop corrections.  While the leading soft
function is known to be free of radiative corrections
\cite{Weinberg:1965nx,Bern:1998sv}, the subleading and sub-subleading operators
do not receive corrections beyond one and two loops respectively, as argued in
\cite{Bern:2014oka,He:2014bga}. Therefore -- upon controlling the soft 
limit of leg $n$ with momentum $\epsilon q^{\mu}$ by the parameter $\epsilon$
-- the $L$-loop graviton amplitude should behave as
\begin{align}
\cM^{\text{$L$-loop}}_{n} \stackrel{\epsilon\to 0}{\longrightarrow} &\quad 
( \frac{1}{\epsilon}\, S^{(0)}_{\GR} +  S^{(1)}_{\GR} + \epsilon \,  S^{(2)}_{\GR})\,
\cM^{\text{$L$-loop}}_{n-1} \\ &  +( S^{(1)\, \text{$1$-loop}}_{\GR} + \epsilon \,  S^{(2)
\text{$1$-loop}}_{\GR})\,
\cM^{\text{$(L-1)$-loop}}_{n-1} + \epsilon \,  S^{(2)
\text{$2$-loop}}_{\GR}\,
\cM^{\text{$(L-2)$-loop}}_{n-1} \, ,
\end{align}
where $\cM_{n}$ denotes the graviton amplitude and $S^{(i)}_{\GR}$ the leading
($i=0$), subleading ($i=1$) or sub-subleading ($i=2$) soft graviton operators
at various loop levels\footnote{In the one-loop case the last term of course
drops out, similarly at tree-level the last two terms are absent.}.

In ref.~\cite{Cachazo:2014dia} it was proposed that loop corrections to soft
theorems could be generally suppressed by taking the soft limit prior to
removing the dimensional regulator. Effectively this amounts to a study of the
soft limit at the level of the loop integrand. However, such an order of limits
appears physically unjustified to us. Moreover, as argued in
refs.~\cite{Bern:2014oka,Bern:2014vva} and extensively shown in
ref.~\cite{Bonocore:2014wua}, this order of limits would invalidate the
cancellation of leading IR divergences when applied to QCD.

\begin{figure}[t]
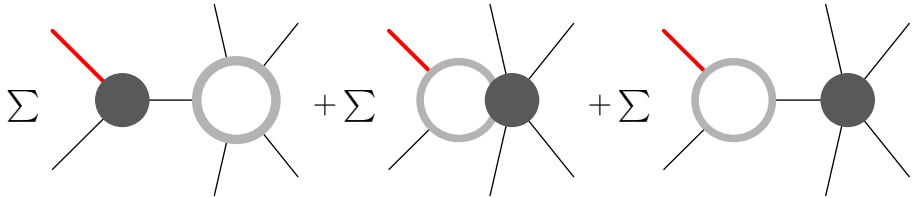

  \begin{center}
    $\sum$
    \raisebox{-1.1cm}{\includegraphics{treeloop.mps}} $+ \sum$
    \raisebox{-1.1cm}{\includegraphics{unfact.mps}} $+ \sum$
    \raisebox{-1.1cm}{\includegraphics{factorized.mps}}
    \caption{One-loop contributions to the soft expansion: the leading soft
    factor acting on the one-loop amplitude, the non-factorizing contributions
    and the factorizing contributions.}
\label{Fig1}
\end{center}
\end{figure}

A central point in the analysis of loop corrections to soft operators concerns
the expected separation into two distinct pieces \cite{Bern:2014vva}: The
``factorizing'' and the ``non-factorizing'' contributions. The factorizing
pieces correspond to the first and the last graph in \figref{Fig1}: while the
leading soft factor acting on the one-loop amplitude is depicted in the first
graph, the last graph shows loop-corrected soft operators acting on the hard
amplitude. The remaining middle graph is associated to the non-factorizing
contributions.

The separation into factorizing and non-factorizing contributions was initially
laid out for collinear limits in \cite{Bern:1995ix}.  Working in dimensional
regularization the poles in $\epsilon$ are to be subtracted from the factorized
diagrams and attributed to the non-factorized contributions to the amplitude.
Related to infrared divergences, the latter are under good control.  This
structure was established in gauge theory in
\cite{Bern:1995ix,Kosower:1999rx,Kosower:2003cz}. 
To our understanding a proof of an analogous separation into factorizing and
non-factorizing pieces does not exist for gravitational amplitudes yet. Given
the results for gauge theory, however, it appears highly plausible to exist at
least at the one-loop level. An analysis of the graviton soft operators
responsible for the leading divergent contributions to the operators
$S^{(i)}_{\GR}$ was performed up to the two-loop level in \cite{Bern:2014oka}.  

Two different concepts are important for the description of soft operators in
the current article: ``locality'' and ``universality''. A soft operator is said
to be local, if the soft particle interacts with each hard particle separately.
Universality of a soft operator implies that the operator is of the same form
for any amplitude, independent of the helicity configuration in question. In
the current article, this universality is implemented by expressing all soft
operators in terms of differential operators in momenta and polarization
vectors.  Accordingly, spinor-helicity expressions appearing in a calculation
of the soft behavior might differ for different helicity configurations
despite originating from a single universal operator expressed in polarization
and momentum vectors.  

Related to the discussion of locality and universality above, an important
subtlety was discussed in ref.~\cite{Bern:2014vva}\footnote{We would like to
thank Zvi Bern, Paolo di Vecchia, Josh Nohle and Scott Davies for pointing out
and explaining this subtlety.}: Contrary to the naive expectation, the
factorizing contributions in gauge theory and gravity do contain
\emph{non-local} parts. While in the
gauge-theory scenario this means that the soft particle interacts with
non-neighboring legs, it implies interaction with more than one hard leg at a
time for gravitational theories. While those contributions can be determined
for every particular configuration, a general formula reproducing those parts
is not known. Accordingly, the analysis in the current article will be limited
to the universal, local and factorizing contributions.

In the present note we extend our tree-level analysis \cite{Broedel:2014fsa} of
the operators appearing in soft-graviton theorems to loop level. In order to do
so, we will \emph{assume} that the separation into a factorizing and a
non-factorizing part extends to higher loops for gravitons in four dimensions.
Starting again from on-shell gauge invariance, factorization and Poincar\'e
symmetry, we adapt the formalism to the appropriate mass dimensions at
different loop levels and identify the restricted class of operators which can
appear at leading, subleading and sub-subleading order. A crucial additional
assumption is again the ``local'' form of the soft operators explained above.
As our method allows to constrain operators composed of hard polarizations and
momenta only, it is blind to possible non-universal contributions to the
factorized part.

Before moving on to gravity, let us comment on the situation in
four-dimensional gauge theory. Here there are no novel loop-level operators
arising from our analysis because of the dimensionless coupling constant.  This
implies in particular that the new subleading operator identified in
\cite{Broedel:2014fsa} captures the form of all radiative corrections. In fact,
the form reported in eq.~(34) of \cite{Broedel:2014fsa} coincides with the
one-loop contribution quoted in \cite{Bern:2014vva}. The undetermined
coefficient depends on the matter content of the gauge theory in question,
whereas the tensorial form of the subleading soft-gluon operator is universal.
However, it is precisely at this order where one first encounters the
\emph{non-local} parts mentioned above: As was analyzed carefully in section
3.2 of \cite{He:2014bga} the single-minus one-loop $(n+1)$-gluon amplitude with
a soft leg of positive helicity develops a factorized \emph{non-local}
subleading soft pole, whose corresponding operator depends on next-to-nearest
neighboring hard legs of the soft leg.  Unfortunately our formalism is unable
to capture this behavior at present.


\section{Method and previous results}

For completeness we repeat our method and the resulting soft-graviton operators
at tree-level \cite{Broedel:2014fsa}.  We denote four-dimensional graviton
amplitudes by $\cM=\delta^{(4)}(P)\, M$ where $P$ is the total momentum. The
momentum of leg $(n+1)$ is expressed as $\epsilon\, q$, which can be made soft
by sending $\epsilon\to 0$. The remaining hard momenta are labeled $p_{a}$ with
$a=1,\ldots , n$. Moreover, we write the polarization tensor of the soft leg as
$E_{\mu\nu}=E_{\mu}\, E_{\nu}$ and similarly for the hard legs
$(E_{a})_{\mu\nu}=E_{a, \mu}\, E_{a, \nu}$. This splitting is an identity in
four dimensions. In this language, the soft-graviton theorem
\cite{Cachazo:2014fwa} reads
\begin{equation}
  \cM_{n+1}(p_{1},\ldots,p_{n},\epsilon q)
  =\Bigl ( \frac 1 \epsilon \, S^{(0)}_\GR + S^{(1)}_\GR
  + \epsilon\, S^{(2)}_\GR  \Bigr )\, \cM_{n}(p_{1},\ldots,p_{n})
  +\cO(\epsilon^{2})\, ,
  \label{softthm}
\end{equation}
where we have suppressed the polarization dependence of the amplitudes $\cM_{m}$. 
The soft operators $S^{(l)}_\GR$ are in general
differential operators in the hard momenta $p_{a}$ and polarizations $E_{a}$ and
also depend on the soft data $q$ and $E$.

\subsection{Constraints on factorized soft operators}
\label{sect:constrsum}

We recall here the assumptions on the form of the soft operators and the
constraints they must satisfy, both of which were analyzed in detail in
ref.~\cite{Broedel:2014fsa}.

\paragraph{Locality.}%
The only assumption on the form of the soft terms is what we termed 
\emph{locality}. We
assume that the soft operators $S^{(l)}_\GR $ can be expressed as a sum of
local operators, each of which acts on the soft leg and one single hard leg:
\begin{equation}
S^{(l)}_\GR  = \sum_{a=1}^{n} S^{(l)}_{a}(E,q, E_{a}, p_{a}; \partial_{p_{a}},
\partial_{E_{a}})\, .
\end{equation}
Clearly this assumption is justified at tree level.  At loop level, however,
there are contributions which are either non-local or non-universal or even
both, as shown in explicit computations of the IR divergences in
refs.~\cite{Bern:2014vva,He:2014bga}.

 \paragraph{Distributional constraint.}%
The self-consistency of eq.~\eqn{softthm} gives rise to the distributional
constraint~\cite{Broedel:2014fsa}
\begin{equation}
  \Bigl ( \frac 1 \epsilon \, S^{(0)}_\GR + S^{(1)}_\GR
+ \epsilon\, S^{(2)}_\GR  \Bigr )\,\, \delta^{(4)}\Bigl (\sum_{a}p_{a}\Bigr ) -
  \delta^{(4)}\Bigl (\sum_{a}p_{a}+ \se\, q \Bigr ) \, 
  \Bigl ( \frac 1 \epsilon \, S^{(0)}_\GR + S^{(1)}_\GR
+ \epsilon\, S^{(2)}_\GR  \Bigr ) = \,\cO(\epsilon^{2})
  \, .
  \label{eq:conscond}
\end{equation}
The expansion of $\delta^{(4)}(\sum_{a}p_{a}+ \se\, q)$ in $\epsilon$ leads to relations
between the operators $S^{(l)}_\GR$. In particular, the terms in
$S^{(l)}_\GR$ coupling to derivatives $\partial/\partial p_{a}^{\mu}$
become constrained by lower-order soft operators $S^{(l'<l)}_\GR$.

\paragraph{Gauge invariance.}%
Naturally, each soft operator must be gauge invariant. For the soft leg this
implies that $S^{(l)}_{a}$ needs to be invariant under the shift 
$
E_{\mu\nu}\to E_{\mu\nu} +q_{(\mu}\, \Lambda_{\nu)}
$
for $q\cdot\Lambda=0$. This can be achieved by demanding
\begin{equation}
  \label{eq:GaugeqE}
q\cdot \frac{\partial}{\partial E}\, S^{(l)}_\GR \sim 0\, ,
\end{equation}
where the symbol $\sim$ indicates vanishing modulo Poincar\'e generators
\begin{equation}
P^{\mu}=\sum_{a}p_{a}^{\mu}\, , \qquad
J^{\mu\nu}= \sum_{a}p_{a}^{\mu}\, \frac{\partial}{\partial p_{a\, \nu}} +
E_{a}^{\mu}\, \frac{\partial}{\partial E_{a\, \nu}} - \mu \leftrightarrow \nu\, .
\end{equation}
Notice that eq.~\eqref{eq:GaugeqE} is a necessary but not sufficient condition
for gauge invariance, because it corresponds to the choice $\Lambda_{\mu} =
E_{\mu}$.  Similarly, a gauge transformation on the soft leg $a$ may be
represented by the generator
\begin{equation}
W_a :=p_a\cdot \frac{\partial}{\partial E_a} \, .
\end{equation}
As $W_{a}\, \mathcal{M}_{n}=0$ the gauge invariance of $\mathcal{M}_{n+1}$
implies the vanishing of the commutator\footnote{%
  Strictly speaking, this is only a sufficient condition for
  $S_{\mathrm{G}}^{(l)}$ to be gauge invariant. Nevertheless, we will see that
  the results in the following sections vanish upon the substitution $E_{a\mu}
  E_{a\nu} \to \Lambda_{a,(\mu} p_{a,\nu)}$ with $\Lambda_a \cdot p_a =
  0$.}
\begin{equation}
  \label{eq:gaugeinv}
  \Bigl[ W_a, S^{(l)}_{\GR} \Bigr]\sim 0\, ,
\end{equation}
which further constrains the form of the soft operators. Here, the zero in
\eqn{eq:gaugeinv} is modulo Poincar\'e and gauge transformations which
annihilate $\mathcal{M}_{n}$.
In ref.~\cite{Broedel:2014fsa} this constraint was satisfied by considering a
particular combination of differential operators to appear in the soft
theorems:
\begin{equation}
\Lambda^{\mu\nu}_{a}:= p_{a}^{\mu}\, \frac{\partial}{\partial p_{a\, \nu} }+
 E_{a}^{\mu}\, \frac{\partial}{\partial E_{a\, \nu}}\, .
\end{equation}
In the current article, however, we will stay more general and refrain from
using this assumption.

\paragraph{Mass dimensions and loop counting.}%
Finally, we need to consider the correct mass dimensions for the soft
operators. In four-dimensional gravity $S^{(n)}_\GR$ ought to have vanishing
mass dimension. This is very important to keep in mind when considering loop
corrections in four dimensions, as the coupling constant $\kappa$ is
dimensionful. Accordingly, for every loop order an extra factor of
$\kappa^{2}\, p_{a}\cdot q$ appears.

\subsection{Ans\"atze}

Let us start by noting Weinberg's leading soft function
\begin{equation}
\label{S0GRtree}
S^{(0) \tree}_\GR =
    \sum_{a=1}^{n} \frac{E_{\mu\nu}\, p_{a}^{\mu}\, p_{a}^{\nu}}{p_{a} \cdot q} \, .
\end{equation}

We then write down the most general ans\"atze for the sub- and sub-subleading
soft operators compatible with the above constraints. The distributional
constraint eq.~\eqref{eq:conscond} and the form of eq.~\eqref{S0GRtree} require
$S^{(1)}_{\GR}$ to be a differential operator of first order in $p_{a}$ whereas
$S^{(2)}_{\GR}$ should be of second order in $p_{a}$.  Hence, the building
blocks for the ans\"atze are single- and double-derivative operators of the
schematic form
\begin{align}
\label{ansatz}
{\tt SD}(r,s) &:= \sum_{a}\, 
(p_{a}\cdot q)^{-r}\, \left [ \, (V\cdot E)\, (V\cdot E)\, (V^{\mu}\, V^{\nu}\, 
L_{\mu \nu})\, \right ]_{\cO(q^{s})}\, , \qquad \nn\\
{\tt DD}(r,s) &:= \sum_{a}
(p_{a}\cdot q)^{-r}\,\left [ \,  (V\cdot E)\, (V\cdot E)\, (V^{\mu}\, V^{\nu}\, 
L_{\mu \nu})\, (V^{\rho}\, V^{\kappa}\, 
L_{\rho \kappa})\, \right ]_{\cO(q^{s})}  , \qquad
\end{align}
where the tensor $L_{\mu\nu}$ can take the values
\begin{equation}
\label{Lchoises}
L_{\mu\nu} \in \left \{ \, p_{a\, \mu}\, \frac{\partial}{\partial p_{a}^{\nu}} \, ,
 E_{a\, \mu}\, \frac{\partial}{\partial E_{a}^{\nu}}
\, \right \}
\, .
\end{equation}
Moreover, the vector $V$ takes one of the three values
\begin{equation}
V^{\mu} \in \left\{\, p_{a}^{\mu} \, ,\,
 n^{\mu}\, \sqrt{p_{a}\cdot q} \, , \,
 q^{\mu} \, \right \}
 \, ,
\end{equation}
where $n$ is a ``dummy'' index vector waiting to be contracted with itself as in
\begin{equation}
(A\cdot n)\, (B\cdot n) \to A\cdot B \, .
\end{equation}
Obviously, only even powers of $n$ are kept in the ans\"atze of eqs.~\eqn{ansatz}. 
Multiple occurrences of $n$ are allowed and encode all possible contractions
weighted with independent coefficients. In this manner we generate all allowed
terms.

The number $s$ in eq.~\eqn{ansatz} counts the effective power of $q$ in the square
brackets.  Note that $s\in [1,3]$ for ${\tt SD}(r,s)$ and $s\in [1,5] $ for
${\tt DD}(r,s)$.  The allowed terms emerging from eq.~\eqn{ansatz} are severely
reduced by the on-shell conditions $E\cdot E= E\cdot q= E_{a}\cdot E_{a}=
E_{a}\cdot p_{a}=0$.  Thus armed, it is straightforward to identify the
potential single- and double-derivative contributions to the soft operators
$S^{(0)}_\GR$, $S^{(1)}_\GR$, $S^{(2)}_\GR$ at tree and loop level, see
\tabref{table1}.

\begin{table}[t]
  \begin{center}
    \begin{tabular}{c|c|c|c|c}
      & Tree & 1-loop & 2-loop & 3-loop \cr
      \hline
      $S^{(0)}_\GR$ &  ${\tt SD}(2,1)$,  ${\tt DD}(3,2)$ &  ${\tt DD}(2,1)$ & -- & --  \cr 
      $S^{(1)}_\GR$ &  \underline{${\tt SD}(2,2)$},  \underline{${\tt DD}(3,3)$} &   ${\tt SD}(1,1)$, ${\tt DD}(2,2)$ &  ${\tt DD}(1,1)$ & -- \cr 
      $S^{(2)}_\GR$ &  \underline{${\tt SD}(2,3)$},  ${\tt DD}(3,4)$ &   \underline{${\tt SD}(1,2)$}, 
      \underline{${\tt DD}(2,3)$} &  ${\tt SD}(0,1)$, ${\tt DD}(1,2)$  &  ${\tt DD}(0,1)$  \cr 
    \end{tabular}
  \end{center}
  \caption{Potential single- and double-derivative 
    contributions to the soft operators permitted by dimensional analysis, soft scaling and
    loop counting. Only the underlined terms turn out to be non-vanishing -- the others vanish
    by either constraints or explicit computation. There are no higher loop contributions.}
  \label{table1}
\end{table}%

Finally let us comment on the possibility of a local soft operator without any derivative
terms. It needs to have the form (cf.~eq.~\eqn{ansatz})
\begin{equation}
{\tt ND}(r,0)=\sum_{a} \frac{(E\cdot p_{a})^{2}}{(p_{a}\cdot q)^{r}} \, ,
\end{equation}
because the only admissible value for $V$ is $p_{a}$. Imposing gauge invariance
on the soft leg, i.e.~acting with $q\cdot \partial_{E}$, 
immediately restricts the invariant operators to $r=1$, which is the Weinberg soft
factor eq.~\eqn{S0GRtree}. Hence, we conclude that there is no other possible
no-derivative structure for $S^{(l)}_\GR$.

\subsection{Tree level results}
After applying the constraints, the tree level results read
\begin{align}
  S^{(0) \tree}_\GR =& 
    \sum_{a=1}^{n} \frac{E_{\mu\nu}\, p_{a}^{\mu}\, p_{a}^{\nu}}{p_{a} \cdot q}, \\
    \label{S1Gtree}
S^{(1)\tree}_{\text{G}} =& \sum_{a=1}^{n} \frac{(p_{a}\cdot E)
    E_{\rho}q_{\sigma}}{p_{a}\cdot q}\, J_{a}^{\rho\sigma} \nn\\
  & \quad + \tilde c\,\sum_{a=1}^{n}  \Bigl ( \frac{(E\cdot
    p_{a})(E_{a}\cdot q)}{p_{a}\cdot q} - E\cdot E_{a} \Bigr ) \Bigr [
  \frac{p_{a}\cdot E}{p_{a}\cdot q}\, q\cdot \frac{\partial}{\partial E_{a}}
  -E\cdot\frac{\partial}{\partial E_{a}}\, \Bigr ] \, , \\
  & \quad + d_1 \sum_{a=1}^{n} \Bigl ( \frac{(E\cdot
    p_{a})(E_{a}\cdot q)}{p_{a}\cdot q} - E\cdot E_{a} \Bigr )^{2}\,
    \frac{\partial}{\partial E_{a}}\cdot   \frac{\partial}{\partial E_{a}}
    \nn\\
  & \quad + d_2 \sum_{a=1}^{n}\frac{1}{p\cdot q}\Bigl ( \frac{(E\cdot
    p_{a})(E_{a}\cdot q)}{p_{a}\cdot q} - E\cdot E_{a} \Bigr )^{2}\,
    (p_{a}\cdot   \frac{\partial}{\partial E_{a}})\,
     (q\cdot   \frac{\partial}{\partial E_{a}}) \nn\\
S^{(2)\tree}_{\text{G}} =& \frac{1}{2} \sum_{a=1}^n  \frac{1}{q\cdot p_a}
  E^{\lambda\sigma} q^\rho q^\gamma \, J_{a,\rho\sigma} \, J_{a,\gamma\lambda} + \nn \\
 & + c_1 \sum_{a=1}^n  
  \frac{1}{q\cdot p_a}
  \biggl(\frac{(p_a\cdot E)(q\cdot E_a)}{q\cdot p_a} - E\cdot E_{a} \biggr )^{2}\, 
  \biggl(q\cdot\frac{\partial}{\partial  E_a}\biggr)^{2} \\ &
  + \tilde c \sum_{a=1}^n \!
\left[ \frac{(p_a\cdot E)(q\cdot E_a)}{q\cdot p_a} - E\cdot E_{a} \right]\!\!\!
\left[\frac{E\cdot p_{a}}{q\cdot p_{a}}\, q\cdot\frac{\partial}{\partial  E_a} 
  - E\cdot\frac{\partial}{\partial  E_a} \right]\!\!\!
\left[\frac{E_{a}\cdot q}{q\cdot p_{a}}\, q\cdot\frac{\partial}{\partial  E_a}
 + q\cdot\frac{\partial}{\partial  p_a}\right]\nn
\end{align}
with the Lorentz generator density
\begin{equation}
J_{a}^{\rho\sigma}=
\Bigl ( p_{a}^{\rho}\,
  \frac{\partial}{\partial p_{a\, \sigma}} + E_{a}^{\, \rho}\,
  \frac{\partial}{\partial E_{a\, \sigma}} - \rho\leftrightarrow \sigma
  \Bigr )\, 
\end{equation}
and four undetermined coefficients $\tilde{c}, c_1, d_1, d_2$. In
ref.~\cite{Broedel:2014fsa} the constants $\tilde{c}$, ${c}_1$ were claimed to
a priori be different for each hard leg. However, since gravity amplitudes are
invariant under any permutation of the external legs, the above soft factors
are compatible with this additional symmetry only if $\tilde{c}$ and $c_1$
agree for all hard legs. Furthermore, the terms proportional to $d_{1}$ and
$d_{2}$ were not displayed in \cite{Broedel:2014fsa}, because double-derivative
terms for the subleading operator were not considered there.  Finally, notice
that this result is derived without any reference to the precise form of the
amplitudes $\cM_{n+1}$ and $\cM_{n}$. 

The only features used are gauge and Lorentz invariance together with total
momentum conservation (ensured by an overall momentum-conserving delta
function).  The above results apply in particular to hard amplitudes
$\cM_{n}(p_{1},\ldots,p_{n})$ involving scalars or photons along with
gravitons. While for a hard scalar the single derivative $\partial_{E_{a}}$
vanishes, for a hard photon only the double derivative $\partial_{E_{a}}^{2}$
does.

In ref.~\cite{Cachazo:2014fwa}, it was shown that $\tilde c=
d_{1}=d_{2}=c_{1}=0$ for tree-level pure-graviton amplitudes in four
dimensions. 


\section{Soft-graviton operators at loop level}

Let us now study the admissible soft operators at loop level. Considering loops
requires taking the mass dimension of the four-dimensional gravitational
coupling $[\kappa]= [p^{-1}]=-1$ into account. Following the derivation of
\cite{Broedel:2014fsa} outlined above, we find several additional permissible
terms in the ans\"{a}tze. A na\"ive way to produce consistent higher loop
corrections satisfying gauge invariance and obeying the distributional
constraints consists of promoting $S^{(n) \text{ $l$-loop}}_{\GR}$ to
$S^{(n+1)\text{ $(l+1)$-loop}}_{\GR}$ by multiplication with $\kappa^{2}\,
(p_{a}\cdot q)$. However, it is quickly checked that the terms proportional to
$J^{\mu\nu}_{a}$ in eqns.~\eqref{S1Gtree} as well as the promotion of
$S^{(0)}_\GR$ violate gauge invariance.

\subsection{Vanishing of the highest-loop contributions}
As stated above, there are no loop corrections to $S^{(0)}_\GR$
\cite{Weinberg:1965nx,Bern:1998sv}. Employing our method, this can be easily
rederived for the local and universal factorizing
contributions by looking at \tabref{table1}: The potential tree-level
derivative operators ${\tt SD}(2,1)$ and ${\tt DD}(3,2)$ vanish by explicit
computation. The double-derivative candidate for a one-loop contribution to
$S^{(0)}_\GR$ takes the form
\begin{equation}
\label{DD21}
{\tt DD}(2,1)= \frac{\kappa^{2}\, (p_{a}\cdot E)^{2}}{p_{a}\cdot q}\,
L^{\mu}_{\nu}\, L_{\mu\rho}\, p^{\nu}_{a}\, p^{\rho}_{a}\, .
\end{equation}
Upon inserting either form of $L_{\mu\nu}$ in eq.~\eqn{Lchoises} we see that
${\tt DD}(2,1)$ vanishes by virtue of the on-shell relations $E_{a}\cdot
E_{a}=p_{a}\cdot p_{a}= E_{a}\cdot p_{a}=0$. Hence there are no factorizing
loop corrections to the leading soft-graviton behavior. This simple argument
directly applies to the potential two-loop contribution to $S^{(1)}_\GR$ in
${\tt DD}(1,1)$ as well as to the potential three-loop contribution to
$S^{(2)}_\GR$ in ${\tt DD}(0,1)$. Since they differ by one and two factors of
$\kappa^{2}(p_{a}\cdot q)$ from eq.~\eqn{DD21} respectively, they vanish by the
same reasoning.

\subsection{Non-vanishing loop contributions}

The absence of loop corrections to the leading soft operator $S^{(0)}_\GR$ has
an immediate consequence for the possible form of the subleading soft operators
at higher loop orders. The distributional constraint eq.~\eqn{eq:conscond}
applied to the higher-loop contributions entails that the subleading soft
operators $S^{(1)}_\GR$ and $S^{(2)}_\GR$ may not contain momentum derivatives
terms but only derivatives with respect to the hard polarizations $E_{a}$.
This further restricts our ans\"atze and leads to the following results: For
the subleading soft operator $S^{(1)}_\GR$ \emph{all} possible loop
contributions vanish because of incompatibility with the constraints.

Turning to the sub-subleading soft operator $S^{(2)}_\GR$ we find that there
are non-vanishing local and universal factorizing contributions
at one-loop order exclusively. We find
\emph{four} contributions to $S^{(2)\text{ 1-loop }}_\GR$, all of which
originate from the tree-level contributions to $S^{(1)\text{tree }}_\GR$ in
eq.~\eqn{S1Gtree} by multiplying with $\kappa^{2}\, (p_{a}\cdot q)$ as
explained above. Collecting everything, we find
\begin{align}
 S^{(2)\text{ 1-loop }}_{\text{G}} =&  
   \phantom{+}\kappa^{2}\, f_{1} \sum_{a=1}^{n}(p_{a}\cdot q)\, 
   \Bigl ( \frac{(E\cdot
    p_{a})(E_{a}\cdot q)}{p_{a}\cdot q} - E\cdot E_{a} \Bigr ) \Bigr [
  \frac{p_{a}\cdot E}{p_{a}\cdot q} \, q\cdot \frac{\partial}{\partial E_{a}}
  - E\cdot\frac{\partial}{\partial E_{a}}\, \Bigr ] \nn\\
  & + \kappa^{2}\, f_{2} \sum_{a=1}^{n} (p_{a}\cdot q)\, \Bigl ( \frac{(E\cdot
    p_{a})(E_{a}\cdot q)}{p_{a}\cdot q} - E\cdot E_{a} \Bigr )^{2}\,
    \frac{\partial}{\partial E_{a}}\cdot   \frac{\partial}{\partial E_{a}}
    \nn\\
  &+ \kappa^{2}\, f_{3} \sum_{a=1}^{n}\Bigl ( \frac{(E\cdot
    p_{a})(E_{a}\cdot q)}{p_{a}\cdot q} - E\cdot E_{a} \Bigr )^{2}\,
    (p_{a}\cdot   \frac{\partial}{\partial E_{a}})\,
     (q\cdot   \frac{\partial}{\partial E_{a}}) \nn\\ 
        & +\kappa^{2}\, f_{4} \sum_{a=1}^{n}\, 
   \Bigl ( \frac{(E\cdot
    p_{a})(E_{a}\cdot q)}{p_{a}\cdot q} - E\cdot E_{a} \Bigr ) 
    \Bigr [ E_{a}\cdot q \,  p_{a}\cdot \frac{\partial}{\partial E_{a}} - p_{a}\cdot q \,
      E_{a}\cdot \frac{\partial}{\partial E_{a}}
   \Bigr ]\times \, \nn\\ & \phantom{+}  \qquad  \times \Bigr [
  \frac{p_{a}\cdot E}{p_{a}\cdot q} \, q\cdot \frac{\partial}{\partial E_{a}}
  - E\cdot\frac{\partial}{\partial E_{a}}\, \Bigr ] 
     \,.
    \label{S2Gloop}
\end{align}
As a matter of fact, the term proportional to $f_{4}$ is effectively equal to
the term proportional to $f_1$ because $E_{a}\partial_{E_{a}}\cM= 2\cM$ and the
remaining $p_{a}\cdot\partial_{E_{a}}$ acts -- together with the further
differential operator -- as a gauge transformation. Hence we may set
$f_4=0$~\footnote{%
  Notice that the term proportional to $f_4$ already originates at tree-level,
  but can be reabsorbed in other terms in a way similar to what explained
  before. }.

Beyond these three terms \emph{all} local and universal factorizing loop
corrections to the soft operators vanish. In particular there cannot be any
factorizing loop corrections to $S^{(0)}_\GR$ or $S^{(1)}_\GR$. Moreover, the
factorizing loop corrections to $S^{(2)}_\GR$ are one-loop exact as shown
above.


\section{Rewriting polarization derivatives} 

All loop corrections appear in the form of polarization derivatives. Using the
completeness relation \cite{Dixon:1996wi}
\begin{equation}
\eta^{\mu\nu} = -E_{a}^{\mu}\, \bar E_{a}^{\nu} - \bar E_{a}^{\mu}\, E_{a}^{\nu}
+ \frac{p_{a}^{\mu}r^{\nu}+
r^{\mu}p_{a}^{\nu}}{r\cdot p_{a}}\, ,
\label{complrel}
\end{equation}
these can be simplified and expressed in a more convenient form. Here $\bar
E_{a}^{\mu}$ is the polarization of opposite helicity compared to $E_{a}^{\mu}$
and $r^{\mu}$ is the reference vector needed to define the
polarizations\footnote{Notice that $E_a \cdot \bar{E}_a = -1$.}.  Using this
language, the $n$-graviton amplitude can be rewritten as
\begin{equation}
  \cM(p_{a},p_1,\ldots, p_{n-1})= E^{\mu}_{a}E^{\nu}_{a}\, \cM_{\mu\nu}(p_{a},p_1,\ldots, p_{n-1})\, .
\end{equation}
Accordingly, we define the conjugate-helicity amplitude as well as the
effective scalar-graviton amplitude as
\begin{align}
  \overbarM{\cM} &:= \bar E^{\mu}_{a} \bar E^{\nu}_{a}\, \cM_{\mu\nu}(p_{a},p_1,\ldots, p_{n-1})\, ,
\nn\\
\cM_{\circ} &:= \bar E^{\mu}_{a} E^{\nu}_{a}\, \cM_{\mu\nu}(p_{a},p_1,\ldots, p_{n-1})
= -\ft 12 \cM_{\mu}^{\mu} +\frac{1}{r\cdot p_a} p_{a}^{\mu}r^{\nu}\cM_{\mu\nu}
\, .
\end{align}
The last relation follows from inserting the completeness relation
eq.~\eqn{complrel} by virtue of $\cM_{\mu\nu}=\cM_{\nu\mu}$.  Using the above
notation, we find after dropping the arguments of the amplitudes
\begin{align}
  \partial_{E_{a}^{\mu}} \cM&= -2 \bar E_{a}^{\mu}\, \cM - 2 E^{\mu}_{a}\,
  \cM_{\circ} + 2\frac{p^{\mu}_{a}}{p_{a}\cdot r}\, (E_{a}^{\rho}r^{\kappa}\cM_{\rho\kappa})\, , \\
 \partial_{E_{a}^{\mu}}\,\partial_{E_{a}^{\nu}} \cM&= 
 2 E^{\mu}_{a}E^{\nu}_{a}\, \overbarM{\cM} +2\bar E^{\mu}_{a}\bar E^{\nu}_{a}\, \cM +
 4 E_{a}^{(\mu}\bar{E}_{a}^{\nu)}\, \cM_{\circ}- 4 \frac{E_{a}^{(\mu}p_{a}^{\nu)}}{r\cdot p_{a}}\, (\bar E_{a}^{\rho}r^{\kappa}\cM_{\rho\kappa}) \nn\\
 & \phantom{=} 
 - 4 \frac{\bar E_{a}^{(\mu}p_{a}^{\nu)}}{r\cdot p_{a}}\, ( E_{a}^{\rho}r^{\kappa}\cM_{\rho\kappa})
 + 2 \frac{p_{a}^{(\mu}p_{a}^{\nu)}}{(r\cdot p_{a})^{2}}\, (r^{\rho}r^{\kappa}\cM_{\rho\kappa})
 + 4 \frac{p_{a}^{(\mu}r^{\nu)}}{(r\cdot p_{a})^{2}}\, (p_{a}^{\rho}r^{\kappa}\cM_{\rho\kappa})\, .
 \end{align}
The particular combinations appearing in the soft operators can now be
expressed in the following way
\begin{align}
  \Bigr [ \frac{p_{a}\cdot E}{p_{a}\cdot q} \, q\cdot \frac{\partial}{\partial
    E_{a}} - E\cdot\frac{\partial}{\partial E_{a}}\, \Bigr ] \, \cM &=
  -2\, \bar T_{a}\, \cM - 2\, T_{a}\, \cM_{\circ}\, , \\
  \frac{\partial}{\partial E_{a}}\cdot \frac{\partial}{\partial E_{a}} \cM &=
  2\, \cM^{\mu}_{\mu}\, ,
\end{align}
where $T_a$ and $\bar T_a$ are the gauge invariant quantities
\begin{equation}
  T_{a}:= \frac{(E\cdot
    p_{a})(E_{a}\cdot q)}{p_{a}\cdot q} - E\cdot E_{a}\, , \qquad
  \bar T_{a}:= \frac{(E\cdot
    p_{a})(\bar E_{a}\cdot q)}{p_{a}\cdot q} - E\cdot \bar E_{a}\,.
\end{equation}
An important relation is 
\begin{equation}
T_{a}\, \bar T_{a}=0 \, ,
\end{equation}
which is most easily seen by choosing $q$ as the reference vector for $E_{a}$
and $\bar E_{a}$. Then $T_{a}\bar T_{a}=(E\cdot E_{a})\, (E\cdot \bar E_{a})$.
The later expression always vanishes: depending on the soft and hard
helicities either the first or the second factor is zero\footnote{
In spinor helicity language one finds
\begin{equation}
T_{a}=  \begin{cases}
 h_{q}=+2, h_{a}=+2: & =-2 \frac{\bev{qa}}{\vev{qa}} \\
 h_{q}=-2, h_{a}=-2: & =-2 \frac{\vev{qa}}{\bev{qa}}\\
\text{else}:  & =0 \\
\end{cases}\, .
\end{equation}}.
Using these relations one may now simplify the soft-operators
at tree and loop level obtained in the previous sections.

\paragraph{Tree level.}%
For the novel tree-level terms in our soft operators we find at subleading order
\begin{equation}
\Delta S^{(1)\tree}_{\text{G}} \cM
= \tilde c\,\sum_{a=1}^{n} T_{a}\,  \Bigl [
  \frac{p_{a}\cdot E}{p_{a}\cdot q}\, q\cdot \frac{\partial}{\partial E_{a}}
  -E\cdot\frac{\partial}{\partial E_{a}}\, \Bigr ] \, \cM
= 
-2\,\tilde c\,\sum_{a=1}^{n} T_{a}^{2}\, \cM_{\circ}\,.
\end{equation}
The contributions proportional to $d_{1,2}$ are of the same form as the
loop-level contribution to $S^{(2)\text{ 1-loop }}_{\text{G}}$ which will be
discussed below. The remaining undetermined terms of the sub-sublea\-ding soft
operator parametrized by $c_{1}$ and $\tilde c$ read
\begin{align}
\Delta_{1} S^{(2)\tree}_{\text{G}} \cM &=
c_1 \,  \sum_{a=1}^n
  \frac{1}{q\cdot p_a}
  T_{a}^{2}\, \biggl(q\cdot\frac{\partial}{\partial  E_a}\biggr)^{2}\cM=  
  2\, c_1\,  \sum_{a=1}^n
  \frac{1}{q\cdot p_a}
  T_{a}^{2}\, (q^{\mu}\, q^{\nu}\, \cM_{\mu\nu})
   \\
\Delta_{2} S^{(2)\tree}_{\text{G}} \cM &=
\tilde c\, \sum_{a=1}^n T_{a}\,\biggl(\frac{E_{a}\cdot q}{q\cdot p_{a}}\, q\cdot\frac{\partial}{\partial  E_a}
 + q\cdot\frac{\partial}{\partial  p_a}\, \biggr)\, 
  \biggl( \frac{E\cdot p_{a}}{q\cdot p_{a}}\, q\cdot\frac{\partial}{\partial  E_a} 
  - E\cdot\frac{\partial}{\partial  E_a} \biggr)\, \cM \nn\\
  &=
\tilde c\, \sum_{a=1}^n T_{a}\,\biggl(\frac{E_{a}\cdot q}{q\cdot p_{a}}\, q\cdot\frac{\partial}{\partial  E_a}
 + q\cdot\frac{\partial}{\partial  p_a}\, \biggr)\, 
  \biggl( -2 \bar T_{a}\, \cM - 2 T_{a}\, \cM_{\circ} \biggr) \nn\\
&= -2 \tilde c\, \sum_{a=1}^n T_{a}^{2}\, \biggl(\frac{E_{a}\cdot q}{q\cdot p_{a}}\, q\cdot\frac{\partial}{\partial  E_a}
 + q\cdot\frac{\partial}{\partial  p_a}\, \biggr)\, \cM_{\circ} \, ,
\end{align}
where in the last step we used the identities $q\cdot \partial_{E_{a}}
T_{a}=0=q\cdot \partial_{E_{a}} \bar T_{a}$ and
$q\cdot \partial_{p_{a}}T_{a}=0=q\cdot \partial_{p_{a}}\bar T_{a}$ as well as
$T_{a}\bar T_{a}=0$.

\paragraph{Loop level.}%
The loop corrections in eq.~\eqn{S2Gloop} can be rewritten as
\begin{align}
 S^{(2)\text{ 1-loop }}_{\text{G}}\, \cM =& \phantom{+}\kappa^{2}\, f_1 \sum_{a=1}^{n}(p_{a}\cdot q)\, 
   T_{a} \Bigl (-2\bar T_{a}\cM - 2T_{a}
    \cM_{\circ} \Bigr ) 
   + \kappa^{2}\, f_2 \sum_{a=1}^{n} (p_{a}\cdot q)\, T_{a}^{2}\,
    \Bigl ( 2\cM_{\mu}^{\mu}\Bigr ) \nn\\ 
    & +2\kappa^{2} f_3 \sum_{a=1}^{n} T_{a}^{2}\, (p_{a}^{\mu}q^{\nu}\cM_{\mu\nu})
    \nn\\=&  
   -2 \kappa^{2}\, f_1 \sum_{a=1}^{n} \, T_{a}^{2}\, (p_{a}\cdot q)\, \cM_{\circ}
   + 2 \kappa^{2}\, f_2 \sum_{a=1}^{n} (p_{a}\cdot q)\, T_{a}^{2}\,
       \cM_{\mu}^{\mu}\nn\\
      & +2\kappa^{2}f_3 \sum_{a=1}^{n}  T_{a}^{2}\, (p_{a}^{\mu}q^{\nu}\cM_{\mu\nu}) \, ,
\end{align}
which can be further simplified by choosing the reference
momentum to equal $q$. This is possible because $T_{a}$ is gauge
invariant by itself. After all, we arrive at the final result
\begin{equation}
 S^{(2)\text{ 1-loop }}_{\text{G}}\, \cM =
    \kappa^{2} e_{1} \sum_{a=1}^{n}  T_{a}^{2}\, (p_{a}^{\mu}q^{\nu}\cM_{\mu\nu})
 +\kappa^{2}\, e_{2} \sum_{a=1}^{n} (p_{a}\cdot q)\, T_{a}^{2}\,
       \cM_{\mu}^{\mu}\, .
       \label{S2lf}
\end{equation}

\paragraph{Interpretation.}%

The loop correction couples to $p_a^\mu q^\nu\cM_{\mu\nu}$ and
$\cM_{\mu}^{\mu}$ exclusively. Note that the latter correction is precisely the
structure that was uncovered in an explicit computation for a $\phi R^{2}$
higher-derivative gravity theory where $\cM_{\circ}$ is a single-scalar
$(n-1)$-graviton amplitude \cite{Bianchi:2014gla}.  To our knowledge there is
no general theorem stating that the polarization-stripped and traced amplitude
$\cM_{\mu}^{\mu}$ vanishes in pure gravity at loop level. While it does so for
four gravitons at tree level, which can be tested using the explicit results in
ref.~\cite{Sannan:1986tz}, calculations are involved already at the one-loop
level. At least the trace of the four-point integrand does not vanish. 


\section{Discussion}

In this paper we have determined all admissible soft factors for
local and universal factorizing loop corrections in four-dimensional gravity.

We found that only the sub-subleading soft operator receives loop corrections.
It turns out that these corrections are one-loop exact and have two
contributions proportional to $p_a^\mu q^\nu\cM_{\mu\nu}$ and
$\cM_{\mu}^{\mu}$. This is consistent with the results in
ref.~\cite{Bern:2014vva}, where loop corrections arising from virtual scalars
in graviton amplitudes were shown to not contribute to subleading soft factors.  

Hence the local and universal factorizing contributions to the soft graviton
operators do not receive corrections at first and second perturbative order.
This implies that the soft-graviton Ward identities of extended BMS symmetry do
not suffer from anomalies caused by local universal operators in the factorized
sector studied here. They will be affected, however, by the non-factorized
contributions entangled with the infrared divergences. It would be interesting
to understand this sector in detail as well.

While the tensorial form of the loop contributions to the sub-subleading soft
graviton theorems is fixed by our analysis, the undetermined scalar prefactors
will reflect the field content of the theory.  For example, our coupling to
$\cM_{\mu}^{\mu}$ exactly matches the structure found in \cite{Bianchi:2014gla}
for $\phi\, R^{2}$ gravity. 

Finally, let us once more comment on gauge theory in four dimensions where the
coupling constant is dimensionless. Therefore, the loop corrections simply
correspond to the undetermined part of the subleading soft operator
$S^{(1)}_{\YM}$ of section 3 in ref.~ \cite{Broedel:2014fsa}, which in turn
agrees with the one-loop correction to $S^{(1)}_{\YM}$ quoted in
\cite{Bern:2014vva}. The undetermined coefficient again depends on the matter
content of the gauge theory in question. The tensorial form of the soft
operators is, however, universal.


\subsubsection*{Acknowledgments}

We thank Z.~Bern, S.~Davies, B.~Schwab, J.~Nohle, P.~di Vecchia and
C.~Vergu for discussions.  JP thanks the Pauli Center for Theoretical Studies
Z\"urich and the Institute for Theoretical Physics at the ETH Z\"urich for
hospitality and support in the framework of a visiting professorship. The work
of MdL and MR is partially supported by grant no.\ 200021-137616 from the Swiss
National Science Foundation.  MdL was also supported in part by FNU through
grant number DFF--1323--00082.

\bibliographystyle{nb}
\bibliography{soft}

\end{document}